\documentclass[aps,pre,reprint,superscriptaddress,showpacs,preprintnumbers,amsmath,amssymb]{revtex4-1}
\usepackage{graphicx}
\graphicspath{{fig/}}
\usepackage{dcolumn}
\usepackage{bm}
\usepackage{color}
\begin{document}
\preprint{K. Saitoh and H. Mizuno}
\title{Sound damping in frictionless granular materials: The interplay between configurational disorder and inelasticity}
\author{Kuniyasu Saitoh}
\affiliation{Department of Physics, Faculty of Science, Kyoto Sangyo University, Motoyama, Kamigamo, Kita-ku, Kyoto 603-8555, Japan}
%
\author{Hideyuki Mizuno}
\affiliation{Graduate School of Arts and Sciences, University of Tokyo, Tokyo, 3-8-1, Japan}
\date{\today}
\begin{abstract}
We numerically investigate sound damping in a model of granular materials in two dimensions.
We simulate evolution of standing waves in disordered frictionless disks and analyze their damped oscillations by velocity autocorrelation functions and power spectra.
We control the strength of inelastic interactions between the disks in contact to examine the effect of energy dissipation on sound characteristics of disordered systems.
Increasing the strength of inelastic interactions, we find that (i) sound softening vanishes
and (ii) sound attenuation due to configurational disorder,\ i.e.\ the Rayleigh scattering at low frequencies and disorder-induced broadening at high frequencies, is completely dominated by the energy dissipation.
Our findings suggest that sound damping in granular media is determined by the interplay between elastic heterogeneities and inelastic interactions.
\end{abstract}
%
\maketitle
%
\section{Introduction}
\label{sec:intro}
Granular materials consisting of macroscopic particles (their sizes range from few $\mu m$ to $mm$) are ubiquitous in nature
and a better understanding of their mechanical properties is crucial to engineering technology \cite{lemaitre}.
In studies of granular materials,\ e.g.\ geophysics, soil mechanics, and civil engineering, \emph{sound characteristics} of granular materials are especially important,\
e.g.\ for geotechnical investigations and understanding of seismic waves \cite{seismic0} and earthquakes \cite{seismic_wave}.
The sound characteristics are influenced not only by grain-level properties but also by complex structures of grains
such as heterogeneous force-chain networks visible in two-dimensional packings of photoelastic disks \cite{gn2,photo2D}.
Because the grains are macroscopic in size, their motions are not affected by thermal fluctuations \cite{general}.
Instead, they can rotate by friction and dissipate kinetic energy by inelastic interactions \cite{dem}.
In recent years, the sound characteristics of grains on a lattice,\ i.e.\ ``granular crystal" \cite{rot_mode0}, have widely been investigated.
For example, the effect of friction on sound dispersions was studied \cite{rot_mode1,rot_mode2,rot_mode3} on the basis of the Hertz-Mindlin theory of contact forces \cite{mindlin}.
In addition, the effect of inelastic interactions (i.e.\ viscous forces between the grains in contact) on the sound characteristics was explored by the theory of granular crystals \cite{rot_mode4}.

However, granular materials in nature are mostly \emph{disordered} and a little is known about sound in disordered granular media \cite{psheng}.
For instance, speeds of sound in disordered granular materials measured by experiments \cite{sound_prop1,sound_prop3} and numerical simulations \cite{sound_prop2,sound_prop4}
deviate from the prediction by effective medium theory based on the Hertz-Mindlin contact \cite{emt0}.
Recently, various anomalies in acoustic sound in disordered systems (such as amorphous solids and glasses) have been pointed out by physicists:
(i) Because of sound dispersions, sound speeds depend on frequency and those in amorphous solids exhibit characteristic ``dips" at intermediate frequencies \cite{sound1,sound3}.
Such \emph{sound softening} was probed by the Brillouin peak of dynamic structure factors (fitted to the damped harmonic oscillator model)
in vitreous silica \cite{sound5,sound6} and Lennard-Jones glasses \cite{md-sound1}, and is more enhanced by increasing the degree of disorder \cite{sound4}.
In general, the intermediate frequencies, where the sound speeds become minimum, are comparable in size with the so-called boson peak frequency $\Omega^{BP}$ \cite{boson0,boson1,boson2,boson5}.
It was found that the Ioffe-Regel (IR) limit for transverse mode $\Omega_T^{IR}$ is close to the boson peak frequency,
whereas that for longitudinal mode $\Omega_L^{IR}$ is much higher,\ i.e.\ $\Omega_T^{IR}\simeq\Omega^{BP}\ll\Omega_L^{IR}$ \cite{shintani}.
On the other hand, the sound speeds increase with the frequency,\ i.e.\ sound hardening occurs, at high frequencies \cite{sound1,sound3,sound5,sound6,md-sound1,sound4}.
(ii) Scattering attenuation of sound is also characteristic of disordered systems such that inhomogeneous elastic moduli are the basics of the theory for \emph{Rayleigh scattering} \cite{psheng}.
In the theory of Rayleigh scattering, attenuation coefficients scale as $\Gamma_\alpha\sim\Omega^{D+1}$ with spatial dimensions $D$,
where $\alpha=L$ and $T$ for longitudinal and transverse modes, respectively \cite{exp-sound9,exp-sound10,lerner_rayleigh,szamel_rayleigh}
(though the logarithmic correction $\Gamma_\alpha\sim\Omega^{D+1}\ln\Omega$ caused by long-ranged spatial correlations of elastic moduli was recently suggested \cite{sound0,sound7}).
However, as the sound speeds exhibit a crossover from softening to hardening \cite{sound1,sound3,sound5,sound6,md-sound1,sound4},
the scaling of attenuation coefficients changes to \emph{disorder-induced broadening} $\Gamma_\alpha\sim\Omega^2$ at high frequencies.
Note that the origin of the disorder-induced broadening in glasses is structural disorder and thus is independent of temperature
\cite{exp-sound0,exp-sound1,exp-sound2,exp-sound3,exp-sound4,exp-sound5,exp-sound6,exp-sound7,exp-sound8,lerner_broadening}.
The crossover of the attenuation coefficients, as well as the boson peak in vibrational density of states,
can be explained by field-theoretical techniques \cite{eh-theory0,boson3,eh-theory1,eh-theory2} which is based on the idea of elastic heterogeneities \cite{eh-MD0,eh-MD1,eh-MD2}.
Moreover, the measurement of sound characteristics enables us to estimate length scales in disordered systems which diverge at the onset of the jamming transition \cite{boson5}.

Despite the success of the theory of granular crystals \cite{rot_mode0,rot_mode1,rot_mode2,rot_mode3,rot_mode4},
it is still not clear how the grain-level properties,\ i.e.\ friction and inelastic interactions, alter the anomalous sound characteristics of disordered systems
such as the sound softening, hardening, IR limits, Rayleigh scattering, and disorder-induced broadening.
Recently, we have studied the influence of friction on sound in disordered granular media \cite{saitoh14}.
However, the effect of inelastic interactions have not yet been investigated, which is our focus in this work.

In this paper, we study sound characteristics of disordered frictionless granular materials by numerical simulations.
In the following, we introduce our numerical method in Sec.\ \ref{sec:method}, show our results in Sec.\ \ref{sec:result}, and discuss our findings in Sec.\ \ref{sec:disc}.
%
\section{Method}
\label{sec:method}
In this section, we explain how to prepare disordered configurations of frictionless granular disks (Sec.\ \ref{sub:disorder}),
introduce linear equations of motion (Sec.\ \ref{sub:motion}), and show our numerical setup (Sec.\ \ref{sub:standing}).
\subsection{Disordered configurations}
\label{sub:disorder}
We generate disordered configurations of two-dimensional disks by molecular dynamics (MD) simulations.
To avoid crystallization of the system, we randomly distribute a $50:50$ binary mixture of $N$ disks in a $L\times L$ square periodic box.
Different kinds of disks have the same mass $m$ and different diameters, $d_S$ and $d_L=1.4d_S$.
The area fraction of the disks is greater than the jamming transition density,\
i.e.\ $\phi\equiv N(d_L^2+d_S^2)/8L^2=0.9>\phi_J\simeq 0.8433$ \cite{gn1}, so that the system is in a solid phase.
The force between the disks, $i$ and $j$, in contact is modeled as a linear elastic force $f_{ij}=k_n\xi_{ij}$,
where $k_n$ represents the stiffness and $\xi_{ij}>0$ is the overlap between the disks.
Then, we minimize elastic energy of the system $E=\sum_{i>j}k_n\xi_{ij}^2/2$ with the aid of FIRE algorithm \cite{FIRE}.
We stop the energy minimization once the maximum acceleration of the disks becomes less than $10^{-9}d_0/t_0^2$ \cite{rs0,rs1}
with the units $d_0\equiv(d_L+d_S)/2$ (i.e.\ the mean disk diameter) and $t_0\equiv\sqrt{m/k_n}$.
In the following, the disk positions after the energy minimization are designated as $\{\bm{r}_i(0)\}$ ($i=1,\dots,N$).
\subsection{Linear equations of motion}
\label{sub:motion}
We introduce equations of motion of frictionless granular disks.
Because our system (with the disk positions $\{\bm{r}_i(0)\}$) is in mechanical equilibrium,
small displacements of the disks at time $t$, $\bm{u}_i(t)\equiv\bm{r}_i(t)-\bm{r}_i(0)$, can be described by linear equations of motion,
\begin{equation}
m\left|\ddot{q}(t)\right\rangle = -\mathcal{D}\left|q(t)\right\rangle-\mathcal{B}\left|\dot{q}(t)\right\rangle~.
\label{eq:motion}
\end{equation}
Here, $\left|q(t)\right\rangle\equiv(\bm{u}_1(t),\dots,\bm{u}_N(t))^\mathrm{T}$ is a $2N$-dimensional displacement vector.
On the right-hand-side of Eq.\ (\ref{eq:motion}), $\mathcal{D}$ and $\mathcal{B}$ are $2N\times2N$ Hessian \cite{vm0,vm1,vm2,vm3} and \emph{damping matrix} \cite{rl0}, respectively.
As shown in Appendix \ref{app:matrix}, the Hessian consists of second derivatives of the elastic energy $E$,
whereas the damping matrix is given by second derivatives of \emph{dissipation function} \cite{rl0}
\begin{equation}
R(t) = \frac{\eta_n}{2}\sum_{i<j}\left\{\dot{\bm{u}}_{ij}(t)\cdot\bm{n}_{ij}\right\}^2~,
\label{eq:dissipation}
\end{equation}
where $\dot{\bm{u}}_{ij}(t)\equiv\dot{\bm{u}}_i(t)-\dot{\bm{u}}_j(t)$ is the relative velocity
and $\bm{n}_{ij}\equiv(\bm{r}_i-\bm{r}_j)/|\bm{r}_i-\bm{r}_j|$ is a normal unit vector.
In Eq.\ (\ref{eq:dissipation}), $\eta_n$ is the viscosity coefficient which determines a microscopic time scale as $t_d\equiv\eta_n/k_n$.
If we use Eq.\ (\ref{eq:dissipation}), Eq.\ (\ref{eq:motion}) is equivalent to a numerical model of frictionless granular disks \cite{dem},
where the second term on the right-hand-side of Eq.\ (\ref{eq:motion}),\ i.e.\ $-\mathcal{B}\left|\dot{q}(t)\right\rangle$,
corresponds with viscous forces between the disks in contact (see Appendix \ref{sub:vforce}).
\subsection{Standing waves}
\label{sub:standing}
To study sound properties of frictionless granular disks, we simulate standing waves of the displacements.
Employing a similar method as in Refs.\ \cite{sound0,boson5,saitoh14},
we numerically integrate the equations of motion [Eq.\ (\ref{eq:motion})] under periodic boundary conditions,
where the initial velocities are given by sinusoidal standing waves,
\begin{equation}
\dot{\bm{u}}_i(0) = \bm{A}\sin(\bm{k}\cdot\bm{r}_i(0))
\label{eq:initial}
\end{equation}
($i=1,\dots,N$).
We use the amplitude vector $\bm{A}$ parallel to the wave vector $\bm{k}$ for the analysis of \emph{longitudinal} (L) \emph{mode},
while $\bm{A}$ perpendicular to $\bm{k}$ (i.e.\ $\bm{A}\cdot\bm{k}=0$) is used to analyze \emph{transverse} (T) \emph{mode}
\footnote{Due to interlocking of the disks, both the L and T modes are excited by any combinations of $\bm{A}$ and $\bm{k}$ though they are most enhanced by the current setup.}.
Note that Eq.\ (\ref{eq:motion}) describes purely harmonic oscillations of the disks around initial positions $\{\bm{r}_i(0)\}$.
Thus, any anharmonic behavior,\ e.g.\ opening and closing contacts \cite{anharmo0,saitoh10}, is not taken into account in our numerical simulations.

In the following, we scale every length and time by $d_0$ and $t_0$, respectively, and use the magnitude $|\bm{A}|=10^{-3}d_0/t_0$ \cite{saitoh14}.
We also introduce a reduced time as
\begin{equation}
\epsilon\equiv\frac{t_d}{t_0}=\frac{\eta_n}{\sqrt{mk_n}}
\label{eq:eps}
\end{equation}
to control \emph{inelasticity} of the system,\
e.g.\ $\epsilon=0$ means that the system is elastic and conserves total energy, whereas large $\epsilon$ represents strong dissipation of kinetic energy \cite{dem}.
%
\section{Results}
\label{sec:result}
In this section, we study sound damping in frictionless granular disks.
First, we analyze time development of the standing waves (Sec.\ \ref{sub:time})
and examine how the inelasticity affects sound characteristics of disordered systems (Sec.\ \ref{sub:disp}).
%
\subsection{Time development of standing waves}
\label{sub:time}
\begin{figure}
\includegraphics[width=\columnwidth]{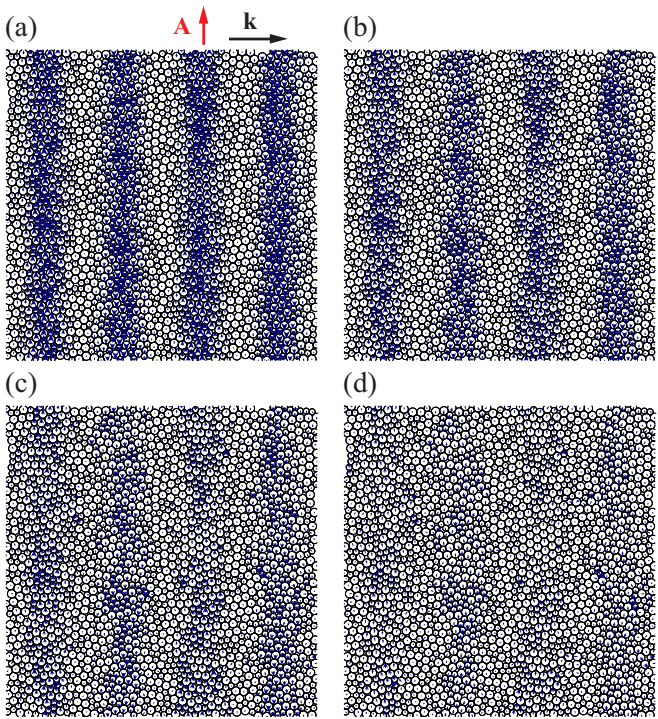}
\caption{(Color online)
Time development of a standing wave, where disk velocities $\dot{\bm{u}}_i(t)$ (arrows) evolve from (a) to (d).
The amplitude vector $\bm{A}$ and wave vector $\bm{k}$ are indicated by the vertical and horizontal arrows in (a), respectively.
The number of disks (circles) is $N=2048$ and we used $k\simeq 0.29d_0^{-1}$ and $\epsilon=0.2$.
\label{fig:demo}}
\end{figure}

By using numerical solutions of Eq.\ (\ref{eq:motion}), we analyze time development of the standing waves.
Figure \ref{fig:demo} displays disk velocities $\dot{\bm{u}}_i(t)$ (arrows), where a small system size with $N=2048$ is used for visualization.
As shown in Fig.\ \ref{fig:demo}(a), the amplitude vector $\bm{A}$ is perpendicular to the wave vector $\bm{k}$, where the wave number is given by $k\equiv|\bm{k}|\simeq 0.29d_0^{-1}$.
We used the inelasticity $\epsilon=0.2$ to demonstrate energy dissipation in the system.
As can be seen, the initial standing wave [Fig.\ \ref{fig:demo}(a)] is attenuated with time [Figs.\ \ref{fig:demo}(b) and (c)] and eventually dies out in a long time limit [Fig.\ \ref{fig:demo}(d)].
We will show that such the attenuation is caused by both \emph{scattering} (due to disordered configurations) and energy dissipation.

In the following analyses, we increase the number of disks and linear system size to $N=2097152$ and $L\simeq 1400d_0$, respectively,
such that we can access the smallest wave number $k_\mathrm{min}=2\pi/L\sim 10^{-3}d_0^{-1}$.
%
\subsubsection{Velocity autocorrelation functions}
\label{sub:auto}
We introduce Fourier transforms of the disk velocities as $\dot{\bm{u}}(\bm{k},t)=\sum_{i=1}^N\dot{\bm{u}}_i(t)e^{-I\bm{k}\cdot\bm{r}_i(t)}$,
where disk positions $\bm{r}_i(t)$ are also obtained from the numerical solutions of Eq.\ (\ref{eq:motion}).
We decompose the Fourier transforms into longitudinal and transverse modes as $\dot{\bm{u}}_L(\bm{k},t)\equiv\{\dot{\bm{u}}(\bm{k},t)\cdot\hat{\bm{k}}\}\hat{\bm{k}}$
and $\dot{\bm{u}}_T(\bm{k},t)\equiv\dot{\bm{u}}(\bm{k},t)-\dot{\bm{u}}_L(\bm{k},t)$, respectively \cite{sound0},
where $\hat{\bm{k}}\equiv\bm{k}/k$ is a unit vector parallel to $\bm{k}$.
Then, we calculate normalized velocity autocorrelation functions (VAFs) as
\begin{equation}
C_\alpha(k,t) = \frac{\langle\dot{\bm{u}}_\alpha(\bm{k},t)\cdot\dot{\bm{u}}_\alpha(-\bm{k},0)\rangle}{\langle|\dot{\bm{u}}_\alpha(\bm{k},0)|^2\rangle}~,
\label{eq:VAF}
\end{equation}
where $\alpha=L$ and $T$ indicate the longitudinal and transverse modes, respectively.

Figure \ref{fig:auto} shows time development of the VAFs (open symbols), where we increase the inelasticity $\epsilon$ as indicated by the arrows.
As can be seen, the oscillations of the L mode [Fig.\ \ref{fig:auto}(a)] are faster than those of the T mode [Fig.\ \ref{fig:auto}(b)] regardless of $\epsilon$.
The amplitudes of VAFs decay in time as sound is attenuated in disordered media \cite{sound0,boson5,saitoh14}.
In addition, the decay is enhanced with the increase of $\epsilon$.
Therefore, sound damping in our system is caused by both the scattering and energy dissipation.
Note that the VAFs are entirely damped without oscillations if the inelasticity is large enough,\ i.e.\ the data for $\epsilon=1$ (triangles).
\subsubsection{Power spectra}
\label{sub:spec}
\begin{figure}
\includegraphics[width=\columnwidth]{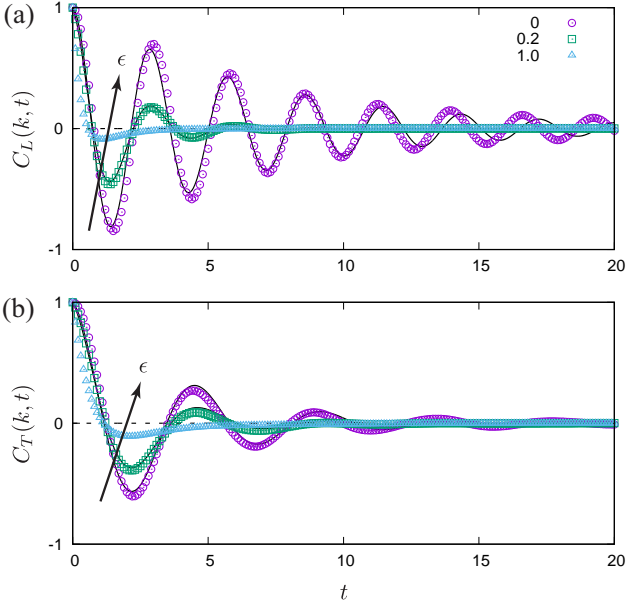}
\caption{(Color online)
Time development of normalized VAFs of the (a) L and (b) T modes, where $k\simeq 2.93d_0^{-1}$.
The inelasticity $\epsilon$ increases as indicated by the arrows and listed in the legend of (a).
The symbols are numerical results of Eq.\ (\ref{eq:VAF}), while the lines represent the damped oscillations [Eq.\ (\ref{eq:damped})] for the date of $\epsilon=0$ and $0.2$.
\label{fig:auto}}
\end{figure}

\begin{figure}
\includegraphics[width=\columnwidth]{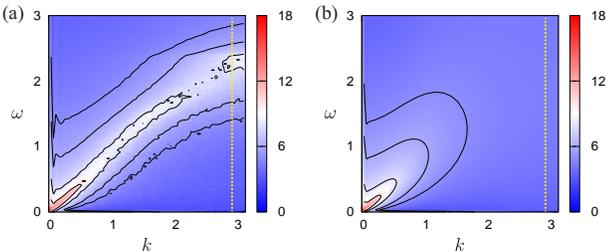}
\caption{(Color online)
Three dimensional plots of logarithm of the power spectrum $\log_{10}S_L(k,\omega)$ (gray scale and contour lines), where (a) $\epsilon=0$ and (b) $1$ are used.
The wave number used in Fig.\ \ref{fig:auto} ($k\simeq 2.93d_0^{-1}$) is indicated by the vertical (yellow) dotted lines.
\label{fig:spec_dyvp}}
\end{figure}

\begin{figure}
\includegraphics[width=\columnwidth]{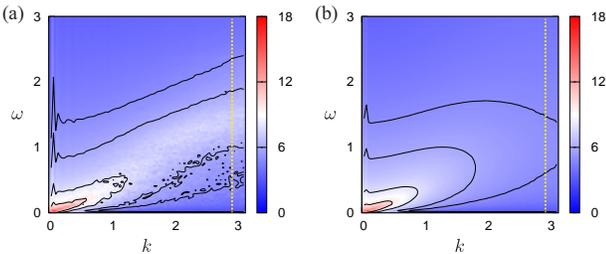}
\caption{(Color online)
Three dimensional plots of $\log_{10}S_T(k,\omega)$ (gray scale and contour lines), where $\epsilon$ and the vertical (yellow) dotted lines are as in Fig.\ \ref{fig:spec_dyvp}.
\label{fig:spec_dyvs}}
\end{figure}

We further investigate the standing waves by power spectra of the L and T modes
\begin{equation}
S_\alpha(k,\omega) = \langle|\tilde{\dot{\bm{u}}}_\alpha(\bm{k},\omega)|^2\rangle~,
\label{eq:spectrum}
\end{equation}
where $\tilde{\dot{\bm{u}}}_\alpha(\bm{k},\omega)\equiv\int_0^\infty\dot{\bm{u}}_\alpha(\bm{k},t)e^{I\omega t}dt$ ($\alpha=L,T$) is the Fourier transform in time.
Figures \ref{fig:spec_dyvp} and \ref{fig:spec_dyvs} display logarithms of the power spectra, $\log_{10}S_\alpha(k,\omega)$,
where \emph{dispersion relation} of the $\alpha$ mode is visible in the region with high intensities (gray scale).
The sound speed defined as the slope of the dispersion relation of the L mode (Fig.\ \ref{fig:spec_dyvp}) is higher than that of the T mode (Fig.\ \ref{fig:spec_dyvs}),
which is consistent with our observations on the VAFs (Fig.\ \ref{fig:auto}).
The power spectra with high wave numbers and frequencies are extremely suppressed
if we introduce the inelasticity $\epsilon=1$ [Figs.\ \ref{fig:spec_dyvp}(b) and \ref{fig:spec_dyvs}(b)].
This means that the energy dissipation is significant for fast oscillations (high $\omega$) at microscopic length scales (high $k$).
Note that the wave number used in Fig.\ \ref{fig:auto} is indicated by the vertical (yellow) dotted lines in Figs.\ \ref{fig:spec_dyvp} and \ref{fig:spec_dyvs}.
The intensities along these lines are infinitesimal if $\epsilon=1$ [Figs.\ \ref{fig:spec_dyvp}(b) and \ref{fig:spec_dyvs}(b)]
such that the VAFs are damped without oscillations (Fig.\ \ref{fig:auto}).
%
\subsection{Sound characteristics}
\label{sub:disp}
To quantitatively extract sound characteristics of the disks from numerical results, we fit a damped oscillation
\begin{equation}
C_\alpha(k,t) = e^{-\Gamma_\alpha(k)t}\cos\Omega_\alpha(k)t
\label{eq:damped}
\end{equation}
to the data of normalized VAF [Eq.\ (\ref{eq:VAF})].
Here, $\Omega_\alpha(k)$ and $\Gamma_\alpha(k)$ correspond to the dispersion relation
and \emph{attenuation coefficient} of the $\alpha$ $(=L,T)$ mode, respectively \cite{sound0,boson5,saitoh14}.
The solid lines in Fig.\ \ref{fig:auto} are the damped oscillations [Eq.\ (\ref{eq:damped})] for $\epsilon=0$ and $0.2$,
where we find perfect agreements with the data by adjusting fitting parameters,\ i.e.\ $\Omega_\alpha(k)$ and $\Gamma_\alpha(k)$, for each $\epsilon$ and $k\simeq 2.93d_0^{-1}$.

\begin{figure}
\includegraphics[width=0.8\columnwidth]{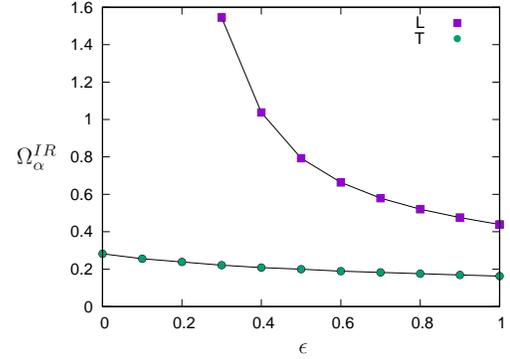}
\caption{(Color online)
The IR limits $\Omega_\alpha^{IR}$ as functions of the inelasticity $\epsilon$,
where the squares and circles are $\alpha=L$ and $T$, respectively (as listed in the legend).
\label{fig:IR-limit}}
\end{figure}

If $\epsilon=1$, however, the oscillations are entirely damped (see triangles in Fig.\ \ref{fig:auto}) so that the VAFs cannot be described by Eq.\ (\ref{eq:damped}).
This means that there is a criterion for measuring sound properties by the damped oscillations.
Therefore, we employ the Ioffe-Regel argument for the criterion, where the damped oscillation [Eq.\ (\ref{eq:damped})] is meaningful only if the condition
\begin{equation}
\frac{\pi\Gamma_\alpha(k)}{\Omega_\alpha(k)} < 1
\label{eq:fit_limit}
\end{equation}
is satisfied \cite{sound4}.
Since the ratio $\pi\Gamma_\alpha(k)/\Omega_\alpha(k)$ is a monotonically increasing function of the wave number $k$, Eq.\ (\ref{eq:fit_limit}) is equivalent to $k<k_\alpha^{IR}(\epsilon)$.
The limit wave number $k_\alpha^{IR}(\epsilon)$ is defined as $\pi\Gamma_\alpha[k_\alpha^{IR}(\epsilon)]/\Omega_\alpha[k_\alpha^{IR}(\epsilon)]=1$,
where $\Omega_\alpha^{IR}(\epsilon)\equiv\Omega_\alpha[k_\alpha^{IR}(\epsilon)]$ is the so-called \emph{Ioffe-Regel (IR) limit} \cite{sound4}.
As shown in Fig.\ \ref{fig:IR-limit}, the IR limits monotonously decrease with the increase of inelasticity $\epsilon$.
As in the case of amorphous solids \cite{sound4,boson5}, the limit for the T mode is less than that for the L mode,\
i.e.\ $\Omega_T^{IR}(\epsilon)<\Omega_L^{IR}(\epsilon)$, regardless of $\epsilon$.
However, $\Omega_L^{IR}(\epsilon)$ is more sensitive to $\epsilon$ than $\Omega_T^{IR}(\epsilon)$
because the viscous forces are introduced to longitudinal relative motions between the disks in contact (and transverse relative motions are not affected by the viscous forces).
Note that $\Omega_L^{IR}(\epsilon)$ for $\epsilon<0.3$ exceeds the maximum frequency which we can access in simulations.
In addition, the wave number used in Fig.\ \ref{fig:auto} ($k\simeq 2.93d_0^{-1}$) exceeds the limits for $\epsilon=1$,\
i.e.\ $k_L^{IR}(1)\simeq 0.52d_0^{-1}$ and $k_T^{IR}(1)\simeq 0.47d_0^{-1}$.

In the following, we only show the data in $\Omega_\alpha<\Omega_\alpha^{IR}(\epsilon)$
and focus on the effects of inelasticity $\epsilon$ on the anomalous sound characteristics of disordered systems.
%
%
\subsubsection{Sound speeds}
First, we clarify the effect of inelasticity on speeds of sound.
Figures \ref{fig:speed_atten_L}(a) and \ref{fig:speed_atten_T}(a) display parametric plots of phase speeds defined as $v_\alpha(k)\equiv\Omega_\alpha(k)/k$
and the dispersion relations $\Omega_\alpha(k)$ obtained by fitting the damped oscillations [Eq.\ (\ref{eq:damped})] to the normalized VAFs [Eq.\ (\ref{eq:VAF})].
In the continuum limit $\Omega_\alpha\rightarrow 0$, the phase speeds converge to finite values regardless of $\epsilon$ \cite{sound_prop1,sound_prop2,sound_prop3,sound_prop4}
so that the viscous forces between the disks in contact do not affect macroscopic speeds of sound.
If $\epsilon=0$, the system conserves total energy and the phase speeds $v_\alpha$ decrease when the frequencies $\Omega_\alpha$ increase from zero (i.e.\ \emph{sound softening}).
Such the sound softening ends at intermediate frequencies,\
i.e.\ $\Omega_L\simeq 4\times 10^{-2}t_0^{-1}$ [Fig.\ \ref{fig:speed_atten_L}(a)] and $\Omega_T\simeq 3\times 10^{-2}t_0^{-1}$ [Fig.\ \ref{fig:speed_atten_T}(a)].
If we further increase the frequencies, the phase speeds start to increase (i.e.\ \emph{sound hardening}), generating small ``dips" at the intermediate frequencies.
The small dips in the phase speeds are characteristic of (energy conserving) disordered media \cite{sound0,boson5}
and have been considered to be a sign of the boson peak in vibrational density of states \cite{boson0,boson1,boson2,boson3}.
Note that the boson peak can be understood as a result of elastic heterogeneities in disordered systems \cite{eh-theory0,eh-theory1,eh-theory2}.
However, depths of the small dips continuously decrease
if we increase the inelasticity $\epsilon$ from zero [as indicated by the arrows in Figs.\ \ref{fig:speed_atten_L}(a) and \ref{fig:speed_atten_T}(a)].
In the case that the energy dissipation is strong enough,\ e.g.\ $\epsilon=1$, the dips entirely vanish and the phase speeds monotonously increase with the frequencies.
Therefore, it seems that there is competition between the influence of elastic heterogeneities and that of energy dissipation in the sound speeds of granular materials.
\subsubsection{Attenuation coefficients}
\begin{figure}
\includegraphics[width=\columnwidth]{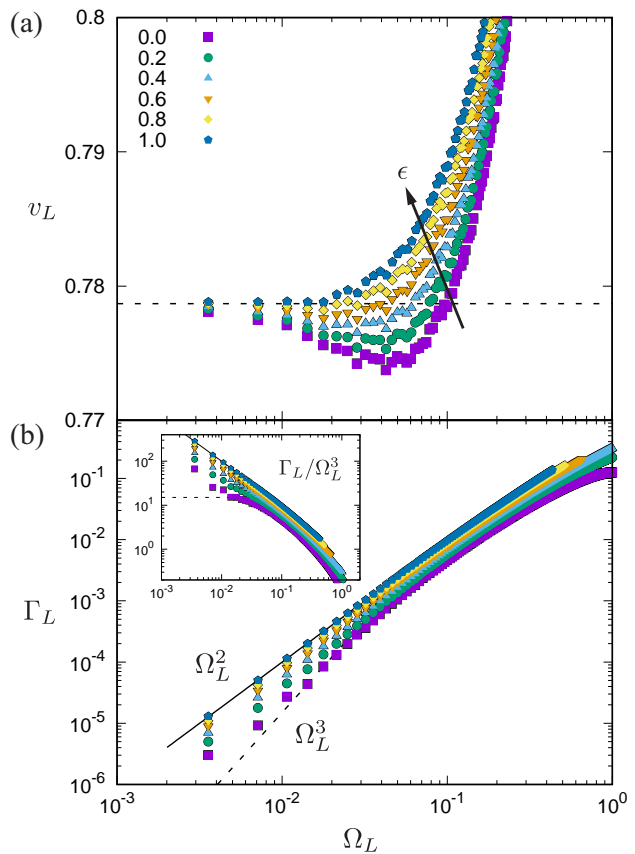}
\caption{(Color online)
Phase speeds $v_L$ and attenuation coefficients $\Gamma_L$ of the L mode (symbols) as functions of the dispersion relations $\Omega_L$.
(a) The inelasticity $\epsilon$ increases as indicated by the arrow and listed in the legend.
The horizontal dotted line indicates the continuum limit $v_L(0)$.
(b) The dotted line represents the Rayleigh scattering (in two dimensions) $\Gamma_L\sim\Omega_L^3$, whereas the solid line is the quadratic growth $\Omega_L^2$.
The inset shows $\Gamma_L/\Omega_L^3$, where the symbols are as in (a).
\label{fig:speed_atten_L}}
\end{figure}

\begin{figure}
\includegraphics[width=\columnwidth]{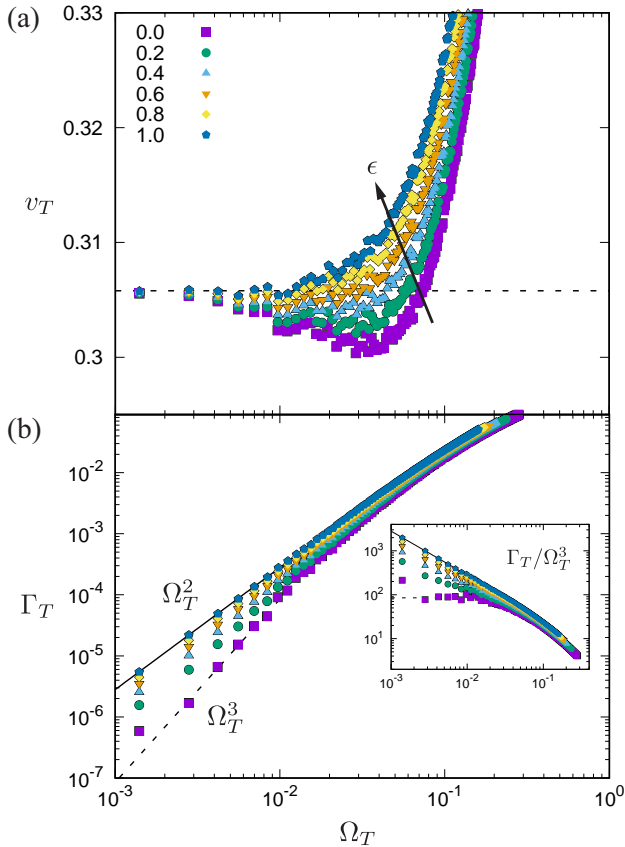}
\caption{(Color online)
Phase speeds $v_T$ and attenuation coefficients $\Gamma_T$ of the T mode (symbols) as functions of the dispersion relations $\Omega_T$.
(a) The inelasticity $\epsilon$ increases as indicated by the arrow and listed in the legend.
The horizontal dotted line indicates the continuum limit $v_T(0)$.
(b) The dotted line represents the Rayleigh scattering $\Gamma_T\sim\Omega_T^3$, whereas the solid line is the quadratic growth $\Omega_T^2$.
The inset shows $\Gamma_T/\Omega_T^3$, where the symbols are as in (a).
\label{fig:speed_atten_T}}
\end{figure}

Next, we focus on the effect on sound attenuation.
Figures \ref{fig:speed_atten_L}(b) and \ref{fig:speed_atten_T}(b) show parametric plots of the attenuation coefficients $\Gamma_\alpha(k)$
and the dispersion relations $\Omega_\alpha(k)$ obtained by fitting Eq.\ (\ref{eq:damped}) to the normalized VAFs [Eq.\ (\ref{eq:VAF})].
In elastic media $\epsilon=0$, the attenuation coefficients exhibit a crossover from the Rayleigh scattering (in two dimensions),\
i.e.\ $\Gamma_\alpha\sim\Omega_\alpha^3$ (dotted lines), to disorder-induced broadening \cite{sound1,sound3,sound4,sound5,sound6,boson5}
around the intermediate frequencies ($\Omega_L\simeq 4\times 10^{-2}t_0^{-1}$ and $\Omega_T\simeq 3\times 10^{-2}t_0^{-1}$),
where we found small dips in the phase speeds [Figs.\ \ref{fig:speed_atten_L}(a) and \ref{fig:speed_atten_T}(a)].
Here, the data for the smallest $\Omega_\alpha$ [squares in Figs.\ \ref{fig:speed_atten_L}(b) and \ref{fig:speed_atten_T}(b)]
deviate from the Rayleigh law (dotted lines) because of the finite size effects \cite{sound0,lerner_broadening,lerner_rayleigh,szamel_rayleigh}.
It is known that the disorder-induced broadening is a structural property of disordered systems and thus insensitive to thermal fluctuations
\cite{exp-sound0,exp-sound1,exp-sound2,exp-sound3,exp-sound4,exp-sound5,exp-sound6,exp-sound7,exp-sound8}.
However, we find that the attenuation coefficients over the whole range of the frequencies continuously increase with the increase of $\epsilon$ [Figs.\ \ref{fig:speed_atten_L}(b) and \ref{fig:speed_atten_T}(b)].
Especially, $\Gamma_\alpha$ below the intermediate frequencies exhibit a continuous change from the Rayleigh scattering to a quadratic growth,\ i.e.\ $\Gamma_\alpha\sim\Omega_\alpha^2$ (solid lines).
Such the quadratic growth in small frequencies, or large length scales, is a general consensus among many macroscopic models of viscoelastic materials \cite{sound2}.
Note that the theory of granular crystals predicts the scaling $\Gamma_\alpha\sim\Omega_\alpha^2$ in the (almost) whole range of the wave number \cite{rot_mode4}.
However, the continuous change from the Rayleigh scattering to quadratic growth is caused by the interplay between disorder and energy dissipation,
and thus cannot be understood by the lattice theories straightforwardly.

%
\section{Discussion}
\label{sec:disc}
In this study, we have investigated sound in frictionless granular media by numerical simulations.
We used disordered configurations of two-dimensional granular disks
so that our study is distinguished from the previous works of ``granular crystals" \cite{rot_mode0,rot_mode1,rot_mode2,rot_mode3,rot_mode4}.
Our focus is how viscous forces between the disks in contact alter sound characteristics of disordered systems
\cite{sound1,sound3,sound5,sound6,md-sound1,sound4,exp-sound0,exp-sound1,exp-sound2,exp-sound3,exp-sound4,exp-sound5,exp-sound6,exp-sound7,exp-sound8}.
Numerically solving the linear equations of motion [Eq.\ (\ref{eq:motion})],
we analyzed time development of the sinusoidal standing waves [Eq.\ (\ref{eq:initial})] by the VAFs [Eq.\ (\ref{eq:VAF})] and power spectra [Eq.\ (\ref{eq:spectrum})].
Damped oscillations of the VAFs showed that the standing waves are attenuated by both the scattering (due to disorder) and energy dissipation.
Moreover, the power spectra in high wave numbers and frequencies are significantly suppressed by the viscous forces, meaning that the energy dissipation takes place at grain-level.
We extracted dispersion relations and attenuation coefficients by fitting the damped oscillations [Eq.\ (\ref{eq:damped})] to our numerical results of the VAFs.
We found that there are limit frequencies,\ i.e.\ the IR limits $\Omega_\alpha^{IR}$, above which the VAFs are entirely damped (without oscillations).
The IR limits are monotonically decreasing functions of the inelasticity $\epsilon$ which quantifies the strength of energy dissipation in the system.
Note that the IR limit for longitudinal mode $\Omega_L^{IR}$ is more sensitive to $\epsilon$ (than that for transverse mode $\Omega_T^{IR}$)
because the viscous forces act on longitudinal relative motions between the disks in contact.
In the continuum limit, phase speeds $v_\alpha$ become independent of $\epsilon$, implying that the viscous forces do not affect macroscopic speeds of sound.
If $\epsilon=0$, the system conserves total energy and the phase speeds exhibit small dips at intermediate frequencies.
This phenomenon known as sound softening is characteristic of disordered systems \cite{sound1,sound3,sound5,sound6,md-sound1,sound4}
and can be related to elastic heterogeneities \cite{eh-theory0,eh-theory1,eh-theory2}.
However, we found that the small dips vanish with the increase of $\epsilon$.
In addition, increasing $\epsilon$, we observed that the attenuation coefficients at low frequencies
exhibit a change from the Rayleigh scattering $\Gamma_\alpha\sim\Omega_\alpha^3$ to a quadratic growth $\Omega_\alpha^2$.
The quadratic growth can be understood by macroscopic models of viscoelastic materials
though the crossover from $\Gamma_\alpha\sim\Omega_\alpha^3$ to $\Omega_\alpha^2$ has never been reported \cite{sound2}.
We also showed that the attenuation coefficients over the whole range of the frequencies increase with the increase of $\epsilon$.
Therefore, the disorder-induced broadening at high frequencies is also affected by inelastic interactions,
which is in sharp contrast to the fact that the disorder-induced broadening is not influenced by temperature
\cite{exp-sound0,exp-sound1,exp-sound2,exp-sound3,exp-sound4,exp-sound5,exp-sound6,exp-sound7,exp-sound8}.

Our main findings are summarized as follows:
Increasing the inelasticity, we find that (i) the sound softening continuously vanishes
and (ii) sound attenuation is enhanced over the whole range of the frequencies,
where the Rayleigh scattering at low frequencies changes to a quadratic growth
and the disorder-induced broadening at high frequencies is dominated by the energy dissipation.
Recently, similar trends have been found in three-dimensional Lennard-Jones glasses \cite{sound_Tdep0,sound_Tdep1,sound_Tdep2},
where the sound softening disappears and the Rayleigh scattering (in three dimensions) $\Gamma_\alpha\sim\Omega_\alpha^4$
changes to $\Omega_\alpha^{3/2}$ with the increase of temperature though the disorder-induced broadening is unchanged.
Because these trends are well predicted by the field-theoretical technique \cite{eh-theory2,sound_Tdep0,sound_Tdep1},
it is an important next step to explore theoretical explanations of our numerical findings (i) and (ii).

In our numerical simulations, we neglected rotational degrees of freedom of the disks.
In reality, however, tangential forces and (the Coulomb or sliding) friction also exist between the disks in contact.
The tangential forces enable the disks to rotate so that, in addition to the longitudinal and transverse modes,
\emph{rotational mode} emerges \cite{rot_mode0,rot_mode1,rot_mode2,rot_mode3,rot_mode4,saitoh14}.
To take into account the rotational degrees of freedom, we need some generalizations of our model, which are left to future work.
Similarly, it is interesting to study how other interaction forces,\
e.g.\ cohesive forces due to capillary bridges in wet granular material \cite{sound_prop2}, affect the sound characteristics.
Moreover, the influence of microstructure \cite{psheng},\ e.g.\ size-distributions and polydispersity, requires more research.
For practical purposes, numerical studies in three dimensions are also important,
where an additional degree of freedom,\ i.e.\ the twisting motion of spheres in contact, enables a \emph{pure rotational mode} \cite{rot_mode1,rot_mode3,rot_mode4}.
In addition, wave diffusion \cite{sound_diff1} and localization phenomena \cite{sound_local1,sound_local2} are other important aspects of sound in granular materials.
%
\begin{acknowledgments}
We thank S. Luding, N.P. Kruyt, X. Jia, H. Steeb, V. Magnanimo, and H. Cheng for fruitful discussions.
This work was financially supported by JSPS KAKENHI Grant Numbers 18K13464, 19K14670, and 20H01868.
\end{acknowledgments}
\appendix*
\section{The Hessian and damping matrix}
\label{app:matrix}
In this appendix, we explain full details of the Hessian (Sec.\ \ref{sub:hessian}) and damping matrix (Sec.\ \ref{sub:damping}).
We also derive viscous forces often used in MD simulations of frictionless granular disks \cite{dem} from the damping matrix (Sec.\ \ref{sub:vforce}).
%
\subsection{The Hessian}
\label{sub:hessian}
The $2N\times2N$ Hessian consists of second derivatives of the elastic energy $E$ with respect to the disk positions $\bm{r}_i=(x_i,y_i)$ as
\begin{equation}
\mathcal{D}=
\begin{pmatrix}
	\frac{\partial^2E}{\partial x_i\partial x_j} & \frac{\partial^2E}{\partial x_i\partial y_j} \\
	\frac{\partial^2E}{\partial y_i\partial x_j} & \frac{\partial^2E}{\partial y_i\partial y_j}
\end{pmatrix}_{i,j=1,\dots,N}~.
\label{eq:hessian}
\end{equation}
The elastic energy is given by the sum of pairwise potentials,\ i.e.\ $E=\sum_{i>j}e_{ij}$ with
\begin{equation}
e_{ij}=\frac{k_n}{2}\xi_{ij}^2~.
\label{eq:eij}
\end{equation}
In Eq.\ (\ref{eq:eij}), $k_n$ is the stiffness and $\xi_{ij}\equiv(d_i+d_j)/2-r_{ij}>0$ represents the overlap between the disks, $i$ and $j$, in contact,
where $d_i$ ($d_j$) is the diameter of the disk $i$ ($j$) and $r_{ij}\equiv|\bm{r}_{ij}|$ with $\bm{r}_{ij}\equiv\bm{r}_i-\bm{r}_j$ is the inter-particle distance between the disks.
The second derivatives of Eq.\ (\ref{eq:eij}) are given by
\begin{eqnarray}
\frac{\partial^2 e_{ij}}{\partial x_i\partial x_i} &=& k_n n_{ijx}^2 - k_n a_{ij} n_{ijy}^2~, \label{ederiv1} \\
\frac{\partial^2 e_{ij}}{\partial x_i\partial y_i} &=& k_n n_{ijx}n_{ijy} + k_n a_{ij} n_{ijx}n_{ijy}~, \label{ederiv2} \\
\frac{\partial^2 e_{ij}}{\partial y_i\partial y_i} &=& k_n n_{ijy}^2 - k_n a_{ij} n_{ijx}^2~, \label{ederiv3}
\end{eqnarray}
where $\bm{n}_{ij}\equiv\bm{r}_{ij}/r_{ij}=(n_{ijx},n_{ijy})$ is the unit vector parallel to the relative position and $a_{ij}\equiv\xi_{ij}/r_{ij}$ is a scaled overlap.
Note that the second derivatives with different indexes ($i$ and $j$) are given by
\begin{equation}
\frac{\partial^2 e_{ij}}{\partial\alpha_i\partial\beta_j} = -\frac{\partial^2 e_{ij}}{\partial\alpha_i\partial\beta_i} \hspace{5mm} (\alpha,\beta=x,y)~.
\label{ederiv4}
\end{equation}
\subsection{Damping matrix}
\label{sub:damping}
The $2N\times2N$ damping matrix consists of second derivatives of the dissipation function $R$ with respect to the disk velocities $\dot{\bm{r}}_i=(\dot{x}_i,\dot{y}_i)$ as
\begin{equation}
\mathcal{B} =
\begin{pmatrix}
	\frac{\partial^2R}{\partial\dot{x}_i\partial\dot{x}_j} & \frac{\partial^2R}{\partial\dot{x}_i\partial\dot{y}_j} \\
	\frac{\partial^2R}{\partial\dot{y}_i\partial\dot{x}_j} & \frac{\partial^2R}{\partial\dot{y}_i\partial\dot{y}_j}
\end{pmatrix}_{i,j=1,\dots,N}~.
\label{eq:dampm}
\end{equation}
The dissipation function defined as Eq.\ (\ref{eq:dissipation}) is the sum of pairwise functions,\ i.e.\ $R=\sum_{i>j}w_{ij}$ with
\begin{equation}
w_{ij} = \frac{\eta_n}{2}\left(\dot{\bm{u}}_{ij}\cdot\bm{n}_{ij}\right)^2~.
\label{eq:wij}
\end{equation}
In Eq.\ (\ref{eq:wij}), $\eta_n$ is the viscosity coefficient and $\dot{\bm{u}}_{ij}=(\dot{u}_{ijx},\dot{u}_{ijy})$ is the relative velocity between the disks in contact.
The second derivatives of Eq.\ (\ref{eq:wij}) are given by
\begin{eqnarray}
\frac{\partial^2 w_{ij}}{\partial\dot{x}_i\partial\dot{x}_i} &=& \eta_n n_{ijx}^2~, \label{wderiv1} \\
\frac{\partial^2 w_{ij}}{\partial\dot{x}_i\partial\dot{y}_i} &=& \eta_n n_{ijx}n_{ijy}~, \label{wderiv2} \\
\frac{\partial^2 w_{ij}}{\partial\dot{y}_i\partial\dot{y}_i} &=& \eta_n n_{ijy}^2~, \label{wderiv3} \\
\frac{\partial^2 w_{ij}}{\partial\alpha_i\partial\beta_j} &=& -\frac{\partial^2 w_{ij}}{\partial\alpha_i\partial\beta_i} \hspace{5mm} (\alpha,\beta=x,y)~. \label{wderiv4}
\end{eqnarray}
\subsection{Viscous forces}
\label{sub:vforce}
Substituting Eq.\ (\ref{eq:dampm}) into the second term on the right-hand-side of Eq.\ (\ref{eq:motion}), we find
\begin{equation}
-\mathcal{B}\left|\dot{q}(t)\right\rangle = \left\{F_{ix}^\mathrm{vis},F_{iy}^\mathrm{vis}\right\}_{i=1,\dots,N}~,
\label{eq:vforce}
\end{equation}
where each component on the right-hand-side is written as
\begin{eqnarray}
F_{ix}^\mathrm{vis} &\equiv& -\sum_j
\Big[\frac{\partial^2 w_{ij}}{\partial\dot{x}_i\partial\dot{x}_i}\dot{u}_{ix}
  +  \frac{\partial^2 w_{ij}}{\partial\dot{x}_i\partial\dot{y}_i}\dot{u}_{iy} \nonumber\\
& & +\frac{\partial^2 w_{ij}}{\partial\dot{x}_i\partial\dot{x}_j}\dot{u}_{jx}
  +  \frac{\partial^2 w_{ij}}{\partial\dot{x}_i\partial\dot{y}_j}\dot{u}_{jy}\Big]~,\label{eq:vforce_x} \\
F_{iy}^\mathrm{vis} &\equiv& -\sum_j
\Big[\frac{\partial^2 w_{ij}}{\partial\dot{y}_i\partial\dot{x}_i}\dot{u}_{ix}
  +  \frac{\partial^2 w_{ij}}{\partial\dot{y}_i\partial\dot{y}_i}\dot{u}_{iy} \nonumber\\
& & +\frac{\partial^2 w_{ij}}{\partial\dot{y}_i\partial\dot{x}_j}\dot{u}_{jx}
  +  \frac{\partial^2 w_{ij}}{\partial\dot{y}_i\partial\dot{y}_j}\dot{u}_{jy}\Big]~.\label{eq:vforce_y}
\end{eqnarray}
From Eqs.\ (\ref{wderiv1})-(\ref{wderiv4}), we find that Eqs.\ (\ref{eq:vforce_x}) and (\ref{eq:vforce_y}) are rewritten as
\begin{eqnarray}
F_{ix}^\mathrm{vis} &=& -\eta_n\sum_j\left(\dot{\bm{u}}_{ij}\cdot\bm{n}_{ij}\right)n_{ijx}~,\label{eq:vforce_x2} \\
F_{iy}^\mathrm{vis} &=& -\eta_n\sum_j\left(\dot{\bm{u}}_{ij}\cdot\bm{n}_{ij}\right)n_{ijy}~,\label{eq:vforce_y2}
\end{eqnarray}
respectively, which correspond with the viscous forces often used in the model of frictionless granular disks \cite{dem}.
Note that the viscous forces [Eqs.\ (\ref{eq:vforce_x2}) and (\ref{eq:vforce_y2})] are also derived from the general expression,
\begin{equation}
\bm{F}_i^\mathrm{vis} = -\frac{\partial R}{\partial\dot{\bm{r}}_i}~.
\label{eq:force_vis}
\end{equation}
\bibliography{sound}
\end{document}